\documentstyle[amssymb,11pt]{article}

\newlength{\minitwocolumn}
\font\teneufm=eufm10
\font\seveneufm=eufm7
\font\fiveeufm=eufm5
\newfam\eufmfam
\textfont\eufmfam=\teneufm
\scriptfont\eufmfam=\seveneufm
\scriptscriptfont\eufmfam=\fiveeufm

\makeatletter
\@addtoreset{equation}{section}
\makeatother

\newtheorem{thm}{Theorem}[section]
\newtheorem{prop}[thm]{Proposition}

\newtheorem{conj}[thm]{Conjecture}

\newtheorem{dfn}[thm]{Definition}
\title{\bf
\huge{\bf
A remark on 
the integrals of motion
associated with level $k$ realization of\\
the elliptic algebra $U_{q,p}(\widehat{sl_2})$}}
\begin{document}
\maketitle
\begin{center}
{
T.KOJIMA$~^\alpha$~~~and~~~J.SHIRAISHI$~^\beta$}
\\~\\
$~^\alpha$~{\it
Department of Mathematics,
College of Science and Technology,
Nihon University,\\
Surugadai, Chiyoda-ku, Tokyo 101-0062, 
JAPAN}\\
$~^\beta$~{\it
Graduate School of Mathematical Science, University of
Tokyo,\\
Komaba, Megro-ku, Tokyo,
153-8914, JAPAN}
\end{center}

~\\

\begin{abstract}
We give one parameter deformation of
level $k$ free field realization of the 
screening current of
the elliptic algebra $U_{q,p}(\widehat{sl_2})$.
By means of these free field realizations,
we construct
infinitly many commutative operators, 
which we call
the nonlocal integrals of motion
associated with
level $k$-realization of 
the elliptic algebra $U_{q,p}(\widehat{sl_2})$.
They are given as integrals 
involving a product of the screening current and 
elliptic theta functions.
This paper give level $k$ generalization of
the nonlocal integrals of motion given in \cite{FKSW}.
\end{abstract}

~\\

\section{Introduction}

One of the results in
V.Bazhanov, S.Lukyanov, Al.Zamolodchikov \cite{BLZ}
is construction of field theoretical analogue
of the commuting transfer matrix ${\bf T}(z)$, acting
on the highest weight representation of the Virasoro algebra.
Their commuting transfer matrix ${\bf T}(z)$ is
the trace of the image of the universal $R$-matrix
associated with the quantum affine symmetry $U_q(\widehat{sl_2})$.
This construction is very simple and
the commutativity $[{\bf T}(z),{\bf T}(w)]=0$ is
direct consequence of the Yang-Baxter equation.
They call the coefficients of the Taylor expansion of ${\bf T}(z)$
the nonlocal integrals of motion.
The higher-rank generalization 
of \cite{BLZ}
is considered in \cite{BHK, Kojima}.
The elliptic deformation of the nonlocal integrals of motion
is considered in \cite{FKSW}.
V.Bazhanov, S.Lukyanov, Al.Zamolodchikov \cite{BLZ}
constructed the continuous transfer matrix ${\bf T}(z)$
by taking the trace of the image of the universal $R$-matrix
associated with $U_q(\widehat{sl_2})$.
However it is not so easy to calculate
the image of the elliptic version of the universal $R$-matrix,
which is obtaind by using the twister \cite{JKOS2}.
Hence the construction method of
the elliptic version  \cite{FKSW} should
be completely different from those in \cite{BLZ}.
Instead of considering the transfer matrix ${\bf T}(z)$,
the authors \cite{FKSW} 
give the integral representation of the integrals of motion 
directly.
The commutativity of the integrals of motion
is not consequence of the Yang-Baxter equation.
It is consequence of the commutative subalgebra
of the Feigin-Odesskii algebra \cite{FO}.
The higher-rank generalization of
\cite{FKSW} is considered in \cite{FKSW2, KS}.
This paper is a continuation of \cite{FKSW, FKSW2, KS}.
This paper give level $k$ generalization of
the nonlocal integrals of motion given in \cite{FKSW}.

The organization of this paper is as following.
In section 2 we give
one parameter ``$s$'' deformation of
the level $k$ free field realization of the 
screening current of
the elliptic algebra $U_{q,p}(\widehat{sl_2})$.
In section 3
we construct
infinitly many commutative operators, 
which are called
the nonlocal integrals of motion
associated with 
the elliptic algebra $U_{q,p}(\widehat{sl_2})$
for level $k$.
In section 3 we state main theorem and
give conjecture.
In appendix we summarize the normal ordering of 
basic operators.

\section{Elliptic current}
In this section
we give one parameter ``$s$'' deformation of
the level $k$ free field realization of the elliptic 
algebra $U_{q,p}(\widehat{sl_2})$.
We fix complex numbers 
$x, r,r^*,s$, $(|x|<1, {\rm Re}(r),{\rm Re}(r^*)>0, 
s \neq 2)$, and $k=r-r^*\neq 0,-2$.
We use symbols
\begin{eqnarray}
~[n]=\frac{x^n-x^{-n}}{x-x^{-1}},~~~[n]_+=x^n+x^{-n}.\nonumber
\end{eqnarray}
We set the parameter $\tau, \tau^*$
\begin{eqnarray}
x=e^{-\pi \sqrt{-1}/r\tau}=e^{-\pi \sqrt{-1}/r^*\tau^*}.
\end{eqnarray}
Let us use parametrization $z=x^{2u}$.
The symbol $[u]_r$
stands for the Jacobi elliptic theta function
\begin{eqnarray}
~[u]_r=x^{\frac{u^2}{r}-u}
\Theta_{x^{2r}}(z),~~
~[u]_{r^*}=x^{\frac{u^2}{r^*}-u}
\Theta_{x^{2r^*}}(z),
\end{eqnarray}
where we have used
\begin{eqnarray}
\Theta_p(z)=(z;p)_\infty (p/z;p)_\infty (p;p)_\infty,~~~
(z;p)_\infty=\prod_{n=0}^\infty (1-p^nz).
\end{eqnarray}
The theta function $[u]_r$ enjoys the quasi-periodicity property
\begin{eqnarray}
[u+r]_r=-[u]_r,~~~[u+r\tau]_r=-e^{-\pi \sqrt{-1}\tau-\frac{2\pi 
\sqrt{-1}}{r}u}[u]_r.
\end{eqnarray}

\subsection{Bosons}

We set
the bosons $\alpha_m^j,\widetilde{\alpha}_m^j, (j=1,2; m\in {\mathbb Z}_{\neq 0})$, 
\begin{eqnarray}
~[\alpha_m^j,\alpha_n^j]&=&-\frac{1}{m}
\frac{[2m][rm]}{[km]
[(r-k)m]}\delta_{m+n,0},~~(j=1,2),\\
~[\alpha_m^1,\alpha_{n}^2]&=&
\frac{1}{m}
\left(
\frac{x^{(-r+k)m}([sm]-[(s-2)m])}{[(r-k)m]}
+
\frac{x^{km}([sm]+[(s-2)m])}{[km]}
\right)\delta_{m+n,0},\\
~[\widetilde{\alpha}_m^j,
\widetilde{\alpha}_n^j]&=&-\frac{1}{m}
\frac{[2m][(r-k)m]}{[km]
[rm]}\delta_{m+n,0},~~(j=1,2),\\
%%%%%%%%%%%%%%%%%%%%%%%%%%%%%%%%%%%%%%%%%%%
%~[\widetilde{\alpha}_m^1,
%\widetilde{\alpha}_n^2]
%&=&\frac{1}{m}\left(
%\frac{[sm][(r-k)m]
%}{[km][rm]}+
%\frac{[(s-2)m][(r-k)m]_+
%}{[km][rm]}
%\right)\delta_{m+n,0},\\
~[\widetilde{\alpha}_m^1,
\widetilde{\alpha}_{n}^2]&=&
\frac{1}{m}
\left(
\frac{x^{rm}(-[sm]+[(s-2)m])}{[rm]}
+
\frac{x^{km}
([sm]+[(s-2)m])}{[km]}
\right)\delta_{m+n,0},\\
~[\alpha_m^j,\widetilde{\alpha}_n^j]&=&
-\frac{1}{m}
\frac{[2m]}{[km]}\delta_{m+n,0},~~(j=1,2),\\
~[\alpha_m^1,\widetilde{\alpha}_n^2]&=&
\frac{1}{m}
\frac{[sm]+[(s-2)m]
}{[km]}\delta_{m+n,0},\\
~[\widetilde{\alpha}_m^1,{\alpha}_n^2]&=&
\frac{1}{m}
\frac{[sm]+[(s-2)m]
}{[km]}\delta_{m+n,0}.
\end{eqnarray}
We set
the bosons $\beta_m^j, \gamma_m^j$, 
$(j=1,2; m \in {\mathbb Z}_{\neq 0})$,
\begin{eqnarray}
~[\beta_m^j,\beta_n^j]&=&
\frac{[2m][(k+2)m]}{m}
\delta_{m+n,0},~(j=1,2),\\
~[\beta_m^1,\beta_n^2]&=&
-\frac{[(k+2)m]
([sm]+[(s-2)m])
}{m}
\delta_{m+n,0},\\
~[\gamma_m^j,\gamma_n^j]&=&
\frac{1}{m}
\frac{[2m]}{[km]}\delta_{m+n,0},~~(j=1,2),\\
~[\gamma_m^1,\gamma_n^2]&=&
-\frac{1}{m}
\frac{[sm]+[(s-2)m]}
{[km]}
\delta_{m+n,0}.
\end{eqnarray}
We set the zero-mode operators $P_0, Q_0$,
$h,\alpha$ and $h_0,h_1,h_2,\alpha_0,\alpha_1,\alpha_2$,
\begin{eqnarray}
&&~[P_0,iQ_0]=1,~[h,\alpha]=2,\\
&&~~[h_0,\alpha_0]=[h_1,\alpha_2]=[h_2,\alpha_1]=(2-s),~~
[h_1,\alpha_1]=[h_2,\alpha_2]=0.
\end{eqnarray}
We set the Fock space ${\cal F}_{K,L}$, $(K,L \in {\mathbb Z})$.
\begin{eqnarray}
&&{\cal F}_{K,L}=
\bigoplus_{n,n_0,n_1,n_2 \in {\mathbb Z}}
{\mathbb C}[\alpha_{-m}^j,\widetilde{\alpha}_{-m}^j,
\beta_{-m}^j, \gamma_{-m}^j,(j=1,2; m \in {\mathbb N}_{\neq 0})
]\otimes 
|K,L\rangle_{n,n_0,n_1,n_2},\nonumber\\
\\
&&|K,L \rangle_{n,n_0,n_1,n_2}=
e^{\left(L\sqrt{\frac{2r}{r-k}}-K
\sqrt{\frac{2(r-k)}{r}}\right)iQ}\otimes
e^{n \alpha}\otimes e^{n_0 \alpha_0}
\otimes e^{n_1 \alpha_1}\otimes e^{n_2 \alpha_2}.
\end{eqnarray}
Upon specialization $s\to 2$,
simplification occures.
\begin{eqnarray}
&&\alpha_m^2=-\alpha_m^1,~~
\widetilde{\alpha}_m^1=\frac{[(r-k)m]}{[rm]}\alpha_m^1,~~
\widetilde{\alpha}_m^2=-\frac{[(r-k)m]}{[rm]}\alpha_m^1,\\
&&\beta_m^2=-\beta_m^1,~~
\gamma_m^2=-\gamma_m^1,~~~
h_0=h_1=h_2=\alpha_0=\alpha_1=\alpha_2=0.
\end{eqnarray}
The bosons $\alpha_m^1,\beta_m^1,\gamma_m^1$ are the same bosons
which were introduced to construct the elliptic current associated with
the elliptic algebra $U_{q,p}(\widehat{sl_2})$ and the deformed Virasoro
algebra $Vir_{q,t}$ \cite{Konno, Matsuo, JKOS}.
In order to construct
infinitly many commutative operators,
we introduce one prameter $s$ deformation of the bosons in 
\cite{Konno, Matsuo, JKOS}.
This additional parameter $s$ plays an important role
in proof of the main theorem.

\subsection{Elliptic current}

We introduce the operators 
$C_j(z), C_j^\dagger(z)$, $(j=1,2)$ acting on
the Fock space ${\cal F}_{J,K}$.
\begin{eqnarray}
C_1(z)&=&e^{-\sqrt{\frac{2r}{k(r-k)}}iQ_0}
e^{-\sqrt{\frac{2r}{k(r-k)}}P_0{\rm log}z}
:\exp\left(-\sum_{m \neq 0}\alpha_{m}^1z^{-m}
\right):,
\\
C_2(z)&=&
e^{\sqrt{\frac{2r}{k(r-k)}}iQ_0}
e^{\sqrt{\frac{2r}{k(r-k)}}P_0{\rm log}z}
:\exp\left(-\sum_{m \neq 0}\alpha_{m}^2z^{-m}\right):,
\\
C_1^\dagger(z)&=&
e^{\sqrt{\frac{2(r-k)}{kr}}iQ_0}
e^{\sqrt{\frac{2(r-k)}{kr}}P_0{\rm log}z}
:\exp\left(\sum_{m \neq 0}
\widetilde{\alpha}_{m}^1z^{-m}\right):,
\\
C_2^\dagger(z)&=&
e^{-\sqrt{\frac{2(r-k)}{kr}}iQ_0}
e^{-\sqrt{\frac{2(r-k)}{kr}}P_0{\rm log}z}
:\exp\left(\sum_{m \neq 0}
\widetilde{\alpha}_{m}^2z^{-m}\right):.
\end{eqnarray}
Here $:*:$ represents normal ordering.
We set the operators $\widetilde{\Psi}_{j,I}(z), 
\widetilde{\Psi}_{j,II}(z), 
\widetilde{\Psi}_{j,I}^\dagger(z),
\widetilde{\Psi}_{j,II}^\dagger(z)
$, $(j=1,2)$ acting on
the Fock space ${\cal F}_{J,K}$.
\begin{eqnarray}
\widetilde{\Psi}_{j,I}(z)
&=&%x^{\frac{1}{2}(-P_1-P_2)}
\exp\left(-(x-x^{-1})
\sum_{m>0}\frac{x^{\frac{km}{2}}}{[m]_+}\beta_{m}^jz^{-m}\right)
\\
&\times&
\exp
\left(-
\sum_{m>0}x^{-\frac{km}{2}}\gamma_{-m}^jz^{m}\right)
\exp
\left(-
\sum_{m>0}x^{\frac{km}{2}}\frac{[(k+1)m]_+}{[m]_+}
\gamma_{m}^jz^{-m}\right),~(j=1,2),\nonumber
\\
\widetilde{\Psi}_{j,II}(z)
&=&%x^{\frac{1}{2}(P_1+P_2)}
\exp\left((x-x^{-1})
\sum_{m>0}
\frac{x^{\frac{km}{2}}}{[m]_+}\beta_{-m}^jz^{m}\right)
\\
&\times&
\exp
\left(-
\sum_{m>0}x^{\frac{km}{2}}
\frac{[(k+1)m]_+}{[m]_+}
\gamma_{-m}^j z^{m}\right)
\exp
\left(-
\sum_{m>0}x^{-\frac{km}{2}}
\gamma_{m}^j
z^{-m}\right),~(j=1,2),\nonumber\\
\widetilde{\Psi}_{j,I}^\dagger(z)
&=&%x^{\frac{1}{2}(P_1-P_2)}
\exp\left((x-x^{-1})
\sum_{m>0}
\frac{x^{-\frac{km}{2}}}{[m]_+}\beta_{m}^j
z^{-m}\right)
\\
&\times&
\exp
\left(
\sum_{m>0}x^{\frac{km}{2}}\gamma_{-m}^j
z^{m}\right)
\exp
\left(
\sum_{m>0}
x^{-\frac{km}{2}}\frac{[(k+1)m]_+}{[m]_+}
\gamma_{m}^jz^{-m}\right),~(j=1,2),\nonumber\\
\widetilde{\Psi}_{j,II}^\dagger(z)
&=&%x^{\frac{1}{2}(-P_1+P_2)}
\exp\left(-(x-x^{-1})
\sum_{m>0}
\frac{x^{-\frac{km}{2}}}{[m]_+}
\beta_{-m}^j
z^{m}\right)
\\
&\times&
\exp
\left(
\sum_{m>0}x^{-\frac{km}{2}}
\frac{[(k+1)m]_+}{[m]_+}
\gamma_{-m}^j
z^{m}\right)
\exp
\left(
\sum_{m>0}x^{\frac{km}{2}}
\gamma_{m}^j
z^{-m}\right),~(j=1,2).\nonumber
\end{eqnarray}
We set the operators $\Psi_{j,I}(z), 
\Psi_{j,II}(z), 
\Psi_{j,I}^\dagger(z),
\Psi_{j,II}^\dagger(z)
$, $(j=1,2)$ acting on
the Fock space ${\cal F}_{J,K}$.
\begin{eqnarray}
\Psi_{1,I}(z)&=&
\widetilde{\Psi}_{1,I}(z)
e^{\alpha+\alpha_0+\alpha_1}
x^{\frac{h}{2}+h_0+h_1}z^{-\frac{h}{k}},\\
\Psi_{1,II}(z)&=&
\widetilde{\Psi}_{1,II}(z)
e^{\alpha+\alpha_0+\alpha_1}
x^{-\frac{h}{2}+h_0-h_1}z^{-\frac{h}{k}},\\
\Psi_{2,I}(z)&=&
\widetilde{\Psi}_{2,I}(z)
e^{-\alpha-\alpha_0+\alpha_2}
x^{-\frac{h}{2}+h_0+h_2}z^{\frac{h}{k}},\\ 
\Psi_{2,II}(z)&=&
\widetilde{\Psi}_{2,II}(z)
e^{-\alpha-\alpha_0+\alpha_2}
x^{\frac{h}{2}+h_0-h_2}z^{\frac{h}{k}},\\
%%%%%%%%%%%%%%%%%%%%%%%%%%%%%%%%%%%%%%%%%%%%%%%%%%
%%%%%%%%%%%%%%%%%%%%%%%%%%%%%%%%%%%%%%%%%%%%%%%%%%
\Psi_{1,I}^\dagger(z)&=&
\widetilde{\Psi}_{1,I}^\dagger(z)
e^{-\alpha-\alpha_0+\alpha_1}
x^{\frac{h}{2}-h_0-h_1}z^{\frac{h}{k}},\\
\Psi_{1,II}^\dagger(z)&=&
\widetilde{\Psi}_{1,II}^\dagger(z)
e^{-\alpha-\alpha_0+\alpha_1}
x^{-\frac{h}{2}-h_0+h_1}z^{\frac{h}{k}},\\
\Psi_{2,I}^\dagger(z)&=&
\widetilde{\Psi}_{2,I}^\dagger(z)
e^{\alpha+\alpha_0+\alpha_2}
x^{-\frac{h}{2}-h_0-h_2}z^{-\frac{h}{k}},\\ 
\Psi_{2,II}^\dagger(z)&=&
\widetilde{\Psi}_{2,II}^\dagger(z)
e^{\alpha+\alpha_0+\alpha_2}
x^{\frac{h}{2}-h_0+h_2}z^{-\frac{h}{k}}.
\end{eqnarray}

\begin{dfn}~~We set the operators 
$E_j(z), F_j(z)$, $(j=1,2)$,
which can be regarded as one parameter
deformation of the level $k$ elliptic currents 
associated with the elliptic algebra $U_{q,p}(\widehat{sl_2})$
\cite{Konno, JKOS}.
\begin{eqnarray}
E_j(z)=C_j(z)\Psi_j(z),~~
F_j(z)=C_j^\dagger(z)\Psi_j^\dagger(z),~~(j=1,2),
\end{eqnarray}
where we have set
\begin{eqnarray}
\Psi_j(z)=\frac{1}{x-x^{-1}}(\Psi_{j,I}(z)-\Psi_{j,II}(z)),~~
\Psi_j^\dagger(z)=\frac{-1}{x-x^{-1}}(\Psi_{j,I}^\dagger(z)-
\Psi_{j,II}^\dagger(z)),~~
(j=1,2).
\end{eqnarray}
\end{dfn}

We have following proposition as direct consequence
of the normal orderings of the basic operators
summarized in appendix.

\begin{prop}~~The elliptic currents $E_j(z)$, $(j=1,2)$ satisfy
the following commutation relations.
\begin{eqnarray}
~&&[u_1-u_2]_{r-k}[u_1-u_2-1]_{r-k}E_j(z_1)E_j(z_2)\nonumber
\\&=&[u_2-u_1]_{r-k}[u_2-u_1-1]_{r-k}E_j(z_2)E_j(z_1),~(j=1,2),
\\
~&&\left[u_1-u_2+\frac{s}{2}\right]_{r-k}
\left[u_1-u_2-\frac{s}{2}+1\right]_{r-k}
E_1(z_1)E_2(z_2)\nonumber\\
&=&
\left[u_2-u_1+\frac{s}{2}\right]_{r-k}
\left[u_2-u_1-\frac{s}{2}+1\right]_{r-k}
E_2(z_2)E_1(z_1).
\end{eqnarray}
The elliptic currents $F_j(z)$, $(j=1,2)$ satisfy
the following commutation relations.
%%%%%%%%%%%%%%%%%%%%%%%%%%%%%%%%%%%%%%%%%%%%%%%%%%%
\begin{eqnarray}
~&&[u_1-u_2]_r[u_1-u_2+1]_rF_j(z_1)F_j(z_2)\nonumber\\
&=&[u_2-u_1]_r[u_2-u_1+1]_rF_j(z_2)F_j(z_1),~(j=1,2),\\
~&&\left[
u_1-u_2-\frac{s}{2}
\right]_r\left[u_1-u_2+\frac{s}{2}-1\right]_r
F_1(z_1)F_2(z_2)\nonumber\\
&=&
\left[u_2-u_1-\frac{s}{2}\right]_r
\left[u_2-u_1+\frac{s}{2}-1\right]_r
F_2(z_2)F_1(z_1).
\end{eqnarray}
%%%%%%%%%%%%%%%%%%%%%%%%%%%%%%%%%%%%%%%%%%%%%%%%%%
The currents $E_j(z)$ and $F_j(z)$ satisfy
\begin{eqnarray}
~[E_j(z_1),F_j(z_2)]&=&\frac{x^{(-1)^j (s-2)}}{x-x^{-1}}
\left(:C_j(z_1)C_j^\dagger(z_2)
\Psi_{j,I}(z_1)\Psi_{j,I}^\dagger(z_2):
\delta\left(\frac{x^kz_2}{z_1}\right)\right.
\\
&-&
\left.
:C_j(z_1)C_j^\dagger(z_2)
\Psi_{j,II}(z_1)\Psi_{j,II}^\dagger(z_2):
\delta\left(\frac{x^{-k}z_2}{z_1}\right)\right),~~
(j=1,2).\nonumber
\end{eqnarray}
Here we have
used the delta-function $\delta(z)=\sum_{n \in {\mathbb Z}}z^n$.
\end{prop}
Upon specialization $s=2$ the currents $E_1(z), F_1(z)$
degenerate to elliptic currents in \cite{JKOS}.
We set
$E_j^{DV}(z)=E_j(z)|_{s=2}$,
$F_j^{DV}(z)=F_j(z)|_{s=2}$, $(j=1,2)$.

\section{Integrals of motion}

In this section we construct
infinitly many commutative operators
${\cal G}_m^*, {\cal G}_m$, $(m \in {\mathbb N})$,
which we call the nonlocal integrals of motion
for level $k$.

\subsection{Nonlocal integrals of motion}

Let us set the theta function $\vartheta_{\alpha}^*(u)$,
$\vartheta_{\alpha}(u)$,
$(\alpha \in {\mathbb C})$ by
\begin{eqnarray}
&&\vartheta^*(u+1)=\vartheta^*(u),~~
\vartheta^*(u+r^*\tau^*)=e^{-2\pi\sqrt{-1}\tau^*-
\frac{2\pi \sqrt{-1}}{r^*}
(2u-\sqrt{\frac{2r r^*}{k}}P_0-\frac{r^*}{k}h)}
\vartheta^*(u),\\
&&\vartheta(u+1)=\vartheta(u),~~
\vartheta(u+r\tau)=e^{-2\pi\sqrt{-1}\tau-
\frac{2\pi \sqrt{-1}}{r}
(2u-\sqrt{\frac{2r r^*}{k}}P_0-\frac{r}{k}h)}
\vartheta(u).
\end{eqnarray}
Let us use the parametrization
$z_j^{(t)}=x^{2u_j^{(t)}}$, $(t=1,2; j=1,2,\cdots,m)$.
\begin{dfn}~~We define the operator
${\cal G}_m^*$ for the regime
${\rm Re}(r)>k$ and $0<{\rm Re}(s)<2$ by
\begin{eqnarray}
{\cal G}_m^*&=&\int \cdots \int_{C^*} \prod_{j=1}^m
\frac{dz_j^{(1)}}{z_j^{(1)}}
\prod_{j=1}^m
\frac{dz_j^{(2)}}{z_j^{(2)}}
E_1(z_1^{(1)})E_1(z_2^{(1)})\cdots E_1(z_m^{(1)})
E_2(z_1^{(2)})E_2(z_2^{(2)})\cdots E_2(z_m^{(2)})\nonumber\\
&\times&\frac{\displaystyle 
\prod_{t=1,2}
\prod_{1\leqq i<j \leqq m}
\left[u_i^{(t)}-u_j^{(t)}\right]_{r-k}
\left[u_j^{(t)}-u_i^{(t)}+1\right]_{r-k}
}{\displaystyle
\prod_{1\leqq i,j \leqq m}
\left[u_i^{(1)}-u_j^{(2)}-\frac{s}{2}\right]_{r-k}
\left[u_j^{(2)}-u_i^{(1)}-\frac{s}{2}+1\right]_{r-k}
}\vartheta^*\left(
\sum_{j=1}^m (u_j^{(2)}-u_j^{(1)})\right),
\end{eqnarray}
were the integral contour $C^*$ encircles $z_j^{(t)}=0$,
$(t=1,2;j=1,2,\cdots,m)$
in such a way that 
$$|z_j^{(t)}|=1,~~(t=1,2; j=1,2,\cdots,m).$$
We define the operator ${\cal G}_m$ for the regime
${\rm Re}(r)>0$ and $0<{\rm Re}(s)<2$ by
\begin{eqnarray}
{\cal G}_m&=&\int \cdots \int_C \prod_{j=1}^m
\frac{dz_j^{(1)}}{z_j^{(1)}}
\prod_{j=1}^m
\frac{dz_j^{(2)}}{z_j^{(2)}}
F_1(z_1^{(1)})F_1(z_2^{(1)})\cdots F_1(z_m^{(1)})
F_2(z_1^{(2)})F_2(z_2^{(2)})\cdots F_2(z_m^{(2)})\nonumber\\
&\times&\frac{\displaystyle 
\prod_{t=1,2}
\prod_{1\leqq i<j \leqq m}
\left[u_i^{(t)}-u_j^{(t)}\right]_{r}
\left[u_j^{(t)}-u_i^{(t)}-1\right]_{r}
}{\displaystyle
\prod_{1\leqq i,j \leqq m}
\left[u_i^{(1)}-u_j^{(2)}+\frac{s}{2}\right]_{r}
\left[u_j^{(2)}-u_i^{(1)}+\frac{s}{2}-1\right]_{r}
}\vartheta\left(
\sum_{j=1}^m (u_j^{(1)}-u_j^{(2)})\right),
\end{eqnarray}
were the integral contour $C^*$ encircles $z_j^{(t)}=0$,
$(t=1,2;j=1,2,\cdots,m)$
in such a way that 
$$|z_j^{(t)}|=1,~~(t=1,2; j=1,2,\cdots,m).$$
We call the operators ${\cal G}_m^*$ and ${\cal G}_m$
the nonlocal integrals of motion for level $k$.
\end{dfn}
The definition of the opeartors
${\cal G}_m^*$, ${\cal G}_m$
for generic $s \in {\mathbb C}, (s \neq 2)$
should be understood as analytic continuation.
In the limit $s \to 2$,
the contour $C^*$, $C$ pinch at $z_j^{(t)}=z_i^{(t')}$.
Hence the definition of ${\cal G}^*_m, {\cal G}_m$
do not hold for $s=2$.
We give modified 
definition of ${\cal G}^*_m, {\cal G}_m$ for $s=2$, below.
We note that
parameter $s\neq 2$ plays an important role in
proof of main theorem \ref{theorem}.

\begin{dfn}~~We define the operator
${\cal G}_m^{DV *}$ for the regime
${\rm Re}(r)>k$ and $s=2$ by
\begin{eqnarray}
{\cal G}_m^{DV *}&=&\int \cdots \int_{C_{Arg}^*} 
\prod_{j=1}^m
\frac{dz_j^{(1)}}{z_j^{(1)}}
\prod_{j=1}^m
\frac{dz_j^{(2)}}{z_j^{(2)}}
E_1^{DV}(z_1^{(1)})\cdots 
E_1^{DV}(z_m^{(1)})
E_2^{DV}(z_1^{(2)})\cdots 
E_2^{DV}(z_m^{(2)})\nonumber\\
&\times&\frac{\displaystyle 
\prod_{t=1,2}
\prod_{1\leqq i<j \leqq m}
\left[u_i^{(t)}-u_j^{(t)}\right]_{r-k}
\left[u_j^{(t)}-u_i^{(t)}+1\right]_{r-k}
}{\displaystyle
\prod_{1\leqq i,j \leqq m}
\left[u_i^{(1)}-u_j^{(2)}-1\right]_{r-k}
\left[u_j^{(2)}-u_i^{(1)}\right]_{r-k}
}\vartheta^*\left(
\sum_{j=1}^m (u_j^{(2)}-u_j^{(1)})\right),
\end{eqnarray}
were the integral contour $C^*_{Arg}$ encircles $z_j^{(t)}=0$,
$(t=1,2;j=1,2,\cdots,m)$
in such a way that 
$$|x^2z^{(2)}_m|,|x^{2r^*}z_m^{(2)}|
<|z_1^{(1)}|<|z_1^{(2)}|
<|z_2^{(1)}|<|z_2^{(2)}|<\cdots
<|z_m^{(1)}|<|z_m^{(2)}|.
$$
We define the operator ${\cal G}_m^{DV}$ for the regime
${\rm Re}(r)>0$ and $s=2$ by
\begin{eqnarray}
{\cal G}_m^{DV}&=&\int \cdots \int_{C_{Arg}} \prod_{j=1}^m
\frac{dz_j^{(1)}}{z_j^{(1)}}
\prod_{j=1}^m
\frac{dz_j^{(2)}}{z_j^{(2)}}
F_1^{DV}(z_1^{(1)})\cdots F_1^{DV}(z_m^{(1)})
F_2^{DV}(z_1^{(2)})\cdots F_2^{DV}(z_m^{(2)})\nonumber\\
&\times&\frac{\displaystyle 
\prod_{t=1,2}
\prod_{1\leqq i<j \leqq m}
\left[u_i^{(t)}-u_j^{(t)}\right]_{r}
\left[u_j^{(t)}-u_i^{(t)}-1\right]_{r}
}{\displaystyle
\prod_{1\leqq i,j \leqq m}
\left[u_i^{(1)}-u_j^{(2)}+1\right]_{r}
\left[u_j^{(2)}-u_i^{(1)}\right]_{r}
}\vartheta\left(
\sum_{j=1}^m (u_j^{(1)}-u_j^{(2)})\right),
\end{eqnarray}
were the integral contour $C_{Arg}$ encircles $z_j^{(t)}=0$,
$(t=1,2;j=1,2,\cdots,m)$
in such a way that 
$$|x^2z^{(2)}_m|,|x^{2r}z_m^{(2)}|
<|z_1^{(1)}|<|z_1^{(2)}|
<|z_2^{(1)}|<|z_2^{(2)}|<\cdots
<|z_m^{(1)}|<|z_m^{(2)}|.
$$
\end{dfn}

\subsection{Main result}

The following is main theorem of this paper.

\begin{thm}~~For the regime $s\neq 2$ and ${\rm Re}(r)>k$,
we have
\begin{eqnarray}
~[{\cal G}_m^*,{\cal G}_n^*]=0,
~~(m,n \in {\mathbb N}).
\end{eqnarray}
For the regime $s\neq 2$ and ${\rm Re}(r)>0$,
we have
\begin{eqnarray}
~[{\cal G}_m,{\cal G}_n]=0,
~~(m,n \in {\mathbb N}).
\end{eqnarray}
\label{theorem}
\end{thm}
We sketch proof of theorem \ref{theorem}. 
Proof is given as the same manner
as level $k=1$ case \cite{FKSW,KS}.
By symmetrization of the screenings $E_j(z)$,
the commutation relation $[{\cal G}_m^*,{\cal G}_n^*]=0$
is reduced to the following sufficient condition
of the theta functions, which is shown by induction
as the same manner as \cite{FKSW,KS}.
We note that this symmetrization procedure
holds only for
$s \neq 2$.
\begin{eqnarray}
&&\sum_{K \cup K^c=\{1,2,\cdots,n+m\}
\atop{|K|=n,|K^c|=m}}
\sum_{L \cup L^c=\{1,2,\cdots,n+m\}
\atop{|L|=n,|L^c|=m}}
\vartheta^*(\sum_{j\in K^c}u_j^{(2)}-\sum_{j \in L^c}u_j^{(1)})
\vartheta^*(\sum_{j\in K}u_j^{(2)}-\sum_{j \in L}u_j^{(1)})
\nonumber\\
&\times&
\prod_{i\in K^c
\atop{p \in K^c}}
\prod_{j\in K^c
\atop{q \in K^c}}
\frac{[u_j^{(2)}-u_p^{(1)}-\frac{s}{2}]_{r-k}
[u_i^{(1)}-u_q^{(2)}-\frac{s}{2}]_{r-k}
[u_p^{(1)}-u_j^{(2)}-\frac{s}{2}+1]_{r-k}
[u_q^{(2)}-u_i^{(1)}-\frac{s}{2}+1]_{r-k}}{
[u_i^{(1)}-u_p^{(1)}]_{r-k}
[u_j^{(2)}-u_q^{(2)}]_{r-k}
[u_p^{(1)}-u_i^{(1)}+1]_{r-k}
[u_q^{(2)}-u_j^{(2)}+1]_{r-k}}\nonumber\\
&=&\sum_{K \cup K^c=\{1,2,\cdots,n+m\}
\atop{|K|=n,|K^c|=m}}
\sum_{L \cup L^c=\{1,2,\cdots,n+m\}
\atop{|L|=n,|L^c|=m}}
\vartheta^*(\sum_{j\in K^c}u_j^{(2)}-\sum_{j \in L^c}u_j^{(1)})
\vartheta^*(\sum_{j\in K}u_j^{(2)}-\sum_{j \in L}u_j^{(1)})
\nonumber\\
&\times&
\prod_{i\in K^c
\atop{p \in K^c}}
\prod_{j\in K^c
\atop{q \in K^c}}
\frac{[u_q^{(2)}-u_i^{(1)}-\frac{s}{2}]_{r-k}
[u_p^{(2)}-u_j^{(1)}-\frac{s}{2}]_{r-k}
[u_i^{(1)}-u_q^{(2)}-\frac{s}{2}+1]_{r-k}
[u_j^{(2)}-u_p^{(1)}-\frac{s}{2}+1]_{r-k}}{
[u_p^{(1)}-u_i^{(1)}]_{r-k}
[u_q^{(2)}-u_j^{(2)}]_{r-k}
[u_i^{(1)}-u_p^{(1)}+1]_{r-k}
[u_q^{(2)}-u_j^{(2)}+1]_{r-k}}.\nonumber\\
\end{eqnarray}

Naively, when we take the limit $s \to 2$, 
it seems that we have
$[{\cal G}_m^{DV*},{\cal G}_n^{DV*}]=0$.
However, very precicely, in order to take 
the limit $s \to 2$,
we have to consider special treatment
which we call ``renormalized'' limit in \cite{FKSW}.
Here we state only conjecture on the operator ${\cal G}_m^{DV*}$.
Theorem \ref{theorem} give a supporting argument of
the following conjecture.

\begin{conj}~~For the regime $s=2$ and ${\rm Re}(r)>k$ we have
\begin{eqnarray}
~[{\cal G}_m^{DV*},{\cal G}_n^{DV*}]=0
~~(m,n \in {\mathbb N}).
\end{eqnarray}
For the regime $s=2$ and ${\rm Re}(r)>0$ we have
\begin{eqnarray}
[{\cal G}_m^{DV},{\cal G}_n^{DV}]=0,
~~(m,n \in {\mathbb N}).
\end{eqnarray}
\end{conj}

In this paper 
we gave one parameter ``$s$'' deformation of
level $k$ free field realization of the 
screening current of
the elliptic algebra $U_{q,p}(\widehat{sl_2})$.
By means of these free field realizations,
we constructed
infinitly many commutative operators, 
which we call
the nonlocal integrals of motion
associated with 
the elliptic algebra $U_{q,p}(\widehat{sl_2})$
for arbitrary level $k\neq 0,-2$.
They are given as integrals 
involving a product of the screening current and 
Jacobi elliptic theta functions.
The construction of the local integrals of motion 
${\cal I}_m$ for arbitrary level $k$ is open problem.
Elliptic deformation of the extended Virasoro algebra 
is needed for this construction.

\section*{Acknowledgements}
We would like to thank the organizing committee
of the X-th International Conference
on Geometry, Integrability and Quantization 
in Sts.Constantine and Elena, Bulgaria.
We would like to thank Professors 
V Bazhanov, P Bouwknegt, A Chervov,
V Gerdjikov, F Goehmann,
K Hasegawa, M Jimbo, A Kluemper, P Kulish,
W-X Ma, V Mangazeev and I Mladenov for their interst in this work.
This work is partly supported by
Grant-in Aid for Young Scientist
{\bf B} (18740092) from JSPS.

\begin{appendix}

\section{Normal Ordering}
In appendix we summarize the normal orderings
of the basic operatrors.
\begin{eqnarray}
C_j(z_1)C_j(z_2)&=&::
z_1^{\frac{2}{r^*}+\frac{2}{k}}
\frac{(x^{-2+2k}z_2/z_1;x^{2r^*})_\infty 
(x^{-2}z_2/z_1;x^{2k})_\infty}{
(x^{2+2k}z_2/z_1;x^{2r^*})_\infty 
(x^2z_2/z_1;x^{2k})_\infty},~(j=1,2),\\
C_1(z_1)C_2(z_2)&=&::
z_1^{-\frac{2}{r^*}-\frac{2}{k}}
\frac{
(x^sz_2/z_1;x^{2r^*})_\infty 
(x^{2-s}z_2/z_1;x^{2r^*})_\infty}{
(x^{-s}z_2/z_1;x^{2r^*})_\infty 
(x^{s-2}z_2/z_1;x^{2r^*})_\infty
}\nonumber\\
&\times&
\frac{
(x^{s+2k}z_2/z_1;x^{2k})_\infty 
(x^{s-2+2k}z_2/z_1;x^{2k})_\infty}{
(x^{-s+2k}z_2/z_1;x^{2k})_\infty 
(x^{2-s+2k}z_2/z_1;x^{2k})_\infty}
,\\
C_2(z_1)C_1(z_2)&=&::
z_1^{-\frac{2}{r^*}-\frac{2}{k}}
\frac{
(x^{s+2r^*}z_2/z_1;x^{2r^*})_\infty 
(x^{2-s+2r^*}z_2/z_1;x^{2r^*})_\infty}{
(x^{-s+2r^*}z_2/z_1;x^{2r^*})_\infty 
(x^{s-2+2r^*}z_2/z_1;x^{2r^*})_\infty
}\nonumber\\
&\times&
\frac{
(x^{s}z_2/z_1;x^{2k})_\infty 
(x^{s-2}z_2/z_1;x^{2k})_\infty}{
(x^{-s}z_2/z_1;x^{2k})_\infty 
(x^{2-s}z_2/z_1;x^{2k})_\infty},\\
C_j^\dagger(z_1)
C_j^\dagger(z_2)
&=&::z_1^{-\frac{2}{r}+\frac{2}{k}}
\frac{(x^{-2+2k}z_1/z_2;x^{2k})_\infty 
(x^{2+2r}z_1/z_2;x^{2r})_\infty}{
(x^{2+2k}z_2/z_1;x^{2k})_\infty 
(x^{-2+2r}z_2/z_1;x^{2r})_\infty},~(j=1,2),\\
C_1^\dagger(z_1)C_2^\dagger(z_2)&=&::
z_1^{\frac{2}{r}-\frac{2}{k}}
\frac{(x^{s+2k}z_2/z_1;x^{2k})_\infty
(x^{s-2+2k}z_2/z_1;x^{2k})_\infty
}{(x^{-s+2k}z_2/z_1;x^{2k})_\infty 
(x^{2-s+2k}z_2/z_1;x^{2k})_\infty}\nonumber\\
&\times&
\frac{(x^{-s+2r}z_2/z_1;x^{2r})_\infty
(x^{s-2+2r}z_2/z_1;x^{2r})_\infty
}{(x^{s+2r}z_2/z_1;x^{2r})_\infty 
(x^{2-s+2r}z_2/z_1;x^{2r})_\infty}
,\\
C_2^\dagger(z_1)C_1^\dagger(z_2)&=&::
z_1^{\frac{2}{r}-\frac{2}{k}}
\frac{(x^{s}z_2/z_1;x^{2k})_\infty
(x^{s-2}z_2/z_1;x^{2k})_\infty
}{(x^{-s}z_2/z_1;x^{2k})_\infty 
(x^{2-s}z_2/z_1;x^{2k})_\infty}\nonumber\\
&\times&
\frac{(x^{-s}z_2/z_1;x^{2r})_\infty
(x^{s-2}z_2/z_1;x^{2r})_\infty
}{(x^{s}z_2/z_1;x^{2r})_\infty 
(x^{2-s}z_2/z_1;x^{2r})_\infty},\\
C_j(z_1)C_j^\dagger(z_2)&=&::
z_1^{-\frac{2}{k}}
\frac{(x^{2+k}z_2/z_1;x^{2k})_\infty}{
(x^{-2+k}z_2/z_1;x^{2k})_\infty},~(j=1,2),\\
C_j^\dagger(z_1)C_j(z_2)&=&::z_1^{-\frac{2}{k}}
\frac{(x^{2+k}z_2/z_1;x^{2k})_\infty}{
(x^{-2+k}z_2/z_1;x^{2k})_\infty},~(j=1,2),\\
%%%%%%%%%%%%%%%%%%%%%%%%%%%%%%%%%%%%%%%%%%%%%%
%%%%%%%%%%%%%%%%%%%%%%%%%%%%%%%%%%%%%%%%%%%%%%
\widetilde{\Psi}_{1,I}(z_1)
\widetilde{\Psi}_{2,I}(z_2)&=&::
\frac{
(x^{-s}z_2/z_1;x^{2k})_\infty 
(x^{2-s+2k}z_2/z_1;x^{2k})_\infty}{
(x^{s+2k}z_2/z_1;x^{2k})_\infty 
(x^{s-2}z_2/z_1;x^{2k})_\infty},\\
\widetilde{\Psi}_{2,I}(z_1)
\widetilde{\Psi}_{1,I}(z_2)&=&::
\frac{
(x^{-s}z_2/z_1;x^{2k})_\infty 
(x^{2-s+2k}z_2/z_1;x^{2k})_\infty}{
(x^{s+2k}z_2/z_1;x^{2k})_\infty 
(x^{s-2}z_2/z_1;x^{2k})_\infty},\\
\widetilde{\Psi}_{1,II}(z_1)
\widetilde{\Psi}_{2,II}(z_2)&=&::
\frac{
(x^{-s}z_2/z_1;x^{2k})_\infty 
(x^{2-s+2k}z_2/z_1;x^{2k})_\infty}{
(x^{s+2k}z_2/z_1;x^{2k})_\infty 
(x^{s-2}z_2/z_1;x^{2k})_\infty},\\
\widetilde{\Psi}_{2,II}(z_1)
\widetilde{\Psi}_{1,II}(z_2)&=&
::
\frac{
(x^{-s}z_2/z_1;x^{2k})_\infty 
(x^{2-s+2k}z_2/z_1;x^{2k})_\infty}{
(x^{s+2k}z_2/z_1;x^{2k})_\infty 
(x^{s-2}z_2/z_1;x^{2k})_\infty},\\
%%%%%%%%%%%%%%%%%%%%%%%%%%%%%%%%%%%%%%%%%%%%%
%%%%%%%%%%%%%%%%%%%%%%%%%%%%%%%%%%%%%%%%%%%%%%
\widetilde{\Psi}_{1,I}^\dagger(z_1)
\widetilde{\Psi}_{2,I}^\dagger(z_2)&=&::
\frac{
(x^{-s}z_2/z_1;x^{2k})_\infty 
(x^{2-s+2k}z_2/z_1;x^{2k})_\infty}{
(x^{s+2k}z_2/z_1;x^{2k})_\infty 
(x^{s-2}z_2/z_1;x^{2k})_\infty},\\
\widetilde{\Psi}_{2,I}^\dagger(z_1)
\widetilde{\Psi}_{1,I}^\dagger(z_2)&=&::
\frac{
(x^{-s}z_2/z_1;x^{2k})_\infty 
(x^{2-s+2k}z_2/z_1;x^{2k})_\infty}{
(x^{s+2k}z_2/z_1;x^{2k})_\infty 
(x^{s-2}z_2/z_1;x^{2k})_\infty},\\
\widetilde{\Psi}_{1,II}^\dagger(z_1)
\widetilde{\Psi}_{2,II}^\dagger(z_2)&=&::
\frac{
(x^{-s}z_2/z_1;x^{2k})_\infty 
(x^{2-s+2k}z_2/z_1;x^{2k})_\infty}{
(x^{s+2k}z_2/z_1;x^{2k})_\infty 
(x^{s-2}z_2/z_1;x^{2k})_\infty},\\
\widetilde{\Psi}_{2,II}^\dagger(z_1)
\widetilde{\Psi}_{1,II}^\dagger(z_2)&=&
::
\frac{
(x^{-s}z_2/z_1;x^{2k})_\infty 
(x^{2-s+2k}z_2/z_1;x^{2k})_\infty}{
(x^{s+2k}z_2/z_1;x^{2k})_\infty 
(x^{s-2}z_2/z_1;x^{2k})_\infty},\\
%%%%%%%%%%%%%%%%%%%%%%%%%%%%%%%%%%%%%%%%%%%%%
%%%%%%%%%%%%%%%%%%%%%%%%%%%%%%%%%%%%%%%%%%%%%
\widetilde{\Psi}_{1,I}(z_1)
\widetilde{\Psi}_{2,II}(z_2)&=&
::
\frac{
(x^{-s+2k}z_2/z_1;x^{2k})_\infty 
(x^{2-s+2k}z_2/z_1;x^{2k})_\infty}{
(x^{s+2k}z_2/z_1;x^{2k})_\infty 
(x^{s-2+2k}z_2/z_1;x^{2k})_\infty},\\
\widetilde{\Psi}_{2,II}(z_1)
\widetilde{\Psi}_{1,I}(z_2)&=&
::
\frac{
(x^{-s}z_2/z_1;x^{2k})_\infty 
(x^{2-s}z_2/z_1;x^{2k})_\infty}{
(x^{s}z_2/z_1;x^{2k})_\infty 
(x^{s-2}z_2/z_1;x^{2k})_\infty},\\
\widetilde{\Psi}_{1,II}(z_1)
\widetilde{\Psi}_{2,I}(z_2)&=&
::
\frac{
(x^{-s}z_2/z_1;x^{2k})_\infty 
(x^{2-s}z_2/z_1;x^{2k})_\infty}{
(x^{s}z_2/z_1;x^{2k})_\infty 
(x^{s-2}z_2/z_1;x^{2k})_\infty},\\
\widetilde{\Psi}_{2,I}(z_1)
\widetilde{\Psi}_{1,II}(z_2)&=&
::
\frac{
(x^{-s+2k}z_2/z_1;x^{2k})_\infty 
(x^{2-s+2k}z_2/z_1;x^{2k})_\infty}{
(x^{s+2k}z_2/z_1;x^{2k})_\infty 
(x^{s-2+2k}z_2/z_1;x^{2k})_\infty},\\
%%%%%%%%%%%%%%%%%%%%%%%%%%%%%%%%%%%%%%%%%%%%%%%%%%%%
\widetilde{\Psi}_{1,I}^\dagger(z_1)
\widetilde{\Psi}_{2,II}^\dagger(z_2)&=&
::
\frac{
(x^{-s}z_2/z_1;x^{2k})_\infty 
(x^{2-s}z_2/z_1;x^{2k})_\infty}{
(x^{s}z_2/z_1;x^{2k})_\infty 
(x^{s-2}z_2/z_1;x^{2k})_\infty},\\
\widetilde{\Psi}_{2,II}^\dagger(z_1)
\widetilde{\Psi}_{1,I}^\dagger(z_2)&=&
::
\frac{
(x^{-s+2k}z_2/z_1;x^{2k})_\infty 
(x^{2-s+2k}z_2/z_1;x^{2k})_\infty}{
(x^{s+2k}z_2/z_1;x^{2k})_\infty 
(x^{s-2+2k}z_2/z_1;x^{2k})_\infty},\\
\widetilde{\Psi}_{1,II}^\dagger(z_1)
\widetilde{\Psi}_{2,I}^\dagger(z_2)&=&
::
\frac{
(x^{-s+2k}z_2/z_1;x^{2k})_\infty 
(x^{2-s+2k}z_2/z_1;x^{2k})_\infty}{
(x^{s+2k}z_2/z_1;x^{2k})_\infty 
(x^{s-2+2k}z_2/z_1;x^{2k})_\infty},\\
\widetilde{\Psi}_{2,I}^\dagger(z_1)
\widetilde{\Psi}_{1,II}^\dagger(z_2)&=&
::
\frac{
(x^{-s}z_2/z_1;x^{2k})_\infty 
(x^{2-s}z_2/z_1;x^{2k})_\infty}{
(x^{s}z_2/z_1;x^{2k})_\infty 
(x^{s-2}z_2/z_1;x^{2k})_\infty},
\\
%%%%%%%%%%%%%%%%%%%%%%%%%%%%%%%%%%%%%%%%%%%%%%
%%%%%%%%%%%%%%%%%%%%%%%%%%%%%%%%%%%%%%%%%%%%%%
\widetilde{\Psi}_{j,I}(z_1)
\widetilde{\Psi}_{j,I}(z_2)&=&::
(1-z_2/z_1)\frac{(x^{2+2k}z_2/z_1;x^{2k})_\infty}{
(x^{-2}z_2/z_1;x^{2k})_\infty},~(j=1,2),\\
\widetilde{\Psi}_{j,II}(z_1)
\widetilde{\Psi}_{j,II}(z_2)&=&::
(1-z_2/z_1)\frac{(x^{2+2k}z_2/z_1;x^{2k})_\infty}{
(x^{-2}z_2/z_1;x^{2k})_\infty},~(j=1,2),\\
\widetilde{\Psi}_{j,I}(z_1)
\widetilde{\Psi}_{j,II}(z_2)&=&::
\frac{(x^{2+2k}z_2/z_1;x^{2k})_\infty}{
(x^{-2+2k}z_2/z_1;x^{2k})_\infty},~(j=1,2),\\
\widetilde{\Psi}_{j,II}(z_1)
\widetilde{\Psi}_{j,I}(z_2)&=&::
\frac{(x^{2}z_2/z_1;x^{2k})_\infty}{
(x^{-2}z_2/z_1;x^{2k})_\infty},~(j=1,2),\\
%%%%%%%%%%%%%%%%%%%%%%%%%%%%%%%%%%%%%%%%%%%%%%%%%%%%%
\widetilde{\Psi}_{j,I}^\dagger(z_1)
\widetilde{\Psi}_{j,I}^\dagger(z_2)&=&::
(1-z_2/z_1)\frac{(x^{2+2k}z_2/z_1;x^{2k})_\infty}{
(x^{-2}z_2/z_1;x^{2k})_\infty},~(j=1,2),\\
\widetilde{\Psi}_{j,II}^\dagger(z_1)
\widetilde{\Psi}_{j,II}^\dagger(z_2)&=&::
(1-z_2/z_1)\frac{(x^{2+2k}z_2/z_1;x^{2k})_\infty}{
(x^{-2}z_2/z_1;x^{2k})_\infty},~(j=1,2),\\
\widetilde{\Psi}_{j,I}^\dagger(z_1)
\widetilde{\Psi}_{j,II}^\dagger(z_2)&=&::
\frac{(x^{2}z_2/z_1;x^{2k})_\infty}{
(x^{-2}z_2/z_1;x^{2k})_\infty},~(j=1,2),\\
\widetilde{\Psi}_{j,II}^\dagger(z_1)
\widetilde{\Psi}_{j,I}^\dagger(z_2)&=&::
\frac{(x^{2+2k}z_2/z_1;x^{2k})_\infty}{
(x^{-2+2k}z_2/z_1;x^{2k})_\infty},~(j=1,2),
\\
%%%%%%%%%%%%%%%%%%%%%%%%%%%%%%%%%%%%%%%%%%%%%%%%%%%%%%%
\widetilde{\Psi}_{j,I}(z_1)
\widetilde{\Psi}_{j,I}^\dagger(z_2)&=&::
\frac{1}{(1-x^kz_2/z_1)}
\frac{(x^{k-2}z_2/z_1;x^{2k})_\infty}{
(x^{3k+2}z_2/z_1;x^{2k})_\infty},~(j=1,2),\\
\widetilde{\Psi}_{j,I}^\dagger(z_1)
\widetilde{\Psi}_{j,I}(z_2)&=&::
\frac{1}{(1-x^{-k}z_2/z_1)}
\frac{(x^{-k-2}z_2/z_1;x^{2k})_\infty}{
(x^{k+2}z_2/z_1;x^{2k})_\infty},~(j=1,2),\\
\widetilde{\Psi}_{j,I}(z_1)
\widetilde{\Psi}_{j,II}^\dagger(z_2)&=&::
\frac{(x^{-2+k}z_2/z_1;x^{2k})_\infty}{(x^{2+k}z_2/z_1;x^{2k})_\infty},~~(j=1,2),\\
\widetilde{\Psi}_{j,II}^\dagger(z_1)
\widetilde{\Psi}_{j,I}(z_2)&=&::
\frac{(x^{-2+k}z_2/z_1;x^{2k})_\infty}{(x^{2+k}z_2/z_1;x^{2k})_\infty},~~(j=1,2),\\
\widetilde{\Psi}_{j,II}(z_1)
\widetilde{\Psi}_{j,I}^\dagger(z_2)&=&::
\frac{(x^{-2+k}z_2/z_1;x^{2k})_\infty}{(x^{2+k}z_2/z_1;x^{2k})_\infty},~~(j=1,2),\\
\widetilde{\Psi}_{j,I}^\dagger(z_1)
\widetilde{\Psi}_{j,II}(z_2)&=&::
\frac{(x^{-2+k}z_2/z_1;x^{2k})_\infty}{(x^{2+k}z_2/z_1;x^{2k})_\infty},~~(j=1,2),\\
\widetilde{\Psi}_{j,II}(z_1)
\widetilde{\Psi}_{j,II}^\dagger(z_2)&=&::
\frac{1}{(1-x^kz_2/z_1)}
\frac{(x^{-k-2}z_2/z_1;x^{2k})_\infty}{
(x^{k+2}z_2/z_1;x^{2k})_\infty},~(j=1,2),
\\
\widetilde{\Psi}_{j,II}^\dagger(z_1)
\widetilde{\Psi}_{j,II}(z_2)&=&::
\frac{1}{(1-x^kz_2/z_1)}
\frac{(x^{k-2}z_2/z_1;x^{2k})_\infty}{
(x^{3k+2}z_2/z_1;x^{2k})_\infty},~(j=1,2).
\end{eqnarray}

\end{appendix}

\end{document}